\documentclass[11pt,onecolumn,amssymb,nofootinbib]{revtex4}
\usepackage{amsmath, amsthm, amscd, amssymb}
\usepackage{graphicx, braket}
\usepackage{bm}
\usepackage{bbm}
\usepackage{epstopdf}

\begin{document}

\title{\bf Dynamical gravastar simulated horizon from the Tolman-Oppenheimer-Volkoff equation initial value problem with relativistic matter } \bigskip

\author{Stephen L. Adler}
\email{adler@ias.edu} \affiliation{Institute for Advanced Study,
Einstein Drive, Princeton, NJ 08540}
\author{Brent Doherty}
\email{bdoherty211@gmail.com}\affiliation{257 Sayre Drive, Princeton, NJ 08540}

\begin{abstract}

We continue the study of ``dynamical gravastars'', constructed by solving the Tolman-Oppenheimer-Volkoff (TOV) equations with relativistic matter, undergoing a phase transition at high pressure to a state with negative energy density. We define the ``simulated horizon'' as the horizon-like structure that appears, and which is a well-defined concept  within the class of static, spherically symmetric metrics.   Since the ``simulated horizon''   occurs at a radius above where the pressure-induced phase transition is postulated to occur, it is solely a property of the TOV equation with relativistic matter, for appropriate small radius initial conditions.  We survey the formation of a simulated   horizon from this point of view.   Rescaling the problem to fixed initial radius, we plot the ``phase diagram'' in the initial pressure--initial mass plane, showing the range of parameters where a simulated horizon dynamically forms. Reformulating the TOV equations in rescaling-invariant form  yields improved numerical results for the ``phase diagram'',  and gives a simplified model for further study consisting of a 2-dimensional autonomous system of first order differential  equations. Our analysis gives a simple,  natural answer to the question of how the radii where the  phase transitions postulated in previous models of exotic compact objects are dynamically determined.
\end{abstract}

\maketitle
\section{Are astrophysical black holes mathematical black holes, or dynamical gravastars?}

 Observations by the EHT \cite{eht} of Sg A* and M87 confirm that each has the expected exterior spacetime geometry of a black hole of mass $M$, with a light sphere at a radius $3M$ lying outside  the Schwarzschild radius of $2M$.\footnote{The radii $2M$ and $3M$ refer specifically to the spherical polar  coordinate system used throughout this paper.  For a detailed description of Schwarzschild black holes in this  coordinate system, and a comparison with the descriptions in other coordinate systems, see Misner, Thorne, and Wheeler \cite{mtw}.} A key question that remains is whether what lies inside the light sphere is  a true mathematical black hole, or a novel  type of relativistic star or ``exotic compact object'' (ECO),  that appears black-hole like from the outside, but has no horizon or interior singularity.  The ECO literature prior to 2019 has been reviewed by Cardoso and Pani \cite{pani}, and a recent review has been given by Mottola \cite{mottola1}.  Both reviews discuss the seminal ``gravastar'' papers of Mazur and Mattola \cite{mazur}, which are based on assuming a pressure jump in the interior equation of state, from a normal matter equation of state to the ``gravity vacuum'' equation of state proposed by Gliner \cite{gliner}, in which the pressure $p$ is minus the density $\rho$. This is the vacuum equation of state associated with a pure cosmological constant or de Sitter universe, and motivates the name {\bf gravastar} = {\bf gra}vity {\bf va}cuum {\bf star}. Related ideas were proposed via a condensed matter analogy in \cite{other}, \cite{other2}, \cite{khlopov1}, \cite{khlopov2}.  Very recently, a multi-layer gravastar has been described by Jampolski and Rezzolla \cite{jamp}, constructed along similar lines as the original gravastar \cite{mazur}.

 In two recent papers \cite{adler1}, \cite{adler2}, one of us (SLA)     has presented  a theory of ``dynamical gravastars'', constructed by following the conventional analysis of relativistic stars given in the classic book of Zeldovich and Novikov \cite{zeld} and in the more recent text of Camenzind \cite{camen}.    This analysis is based on using as input only the Tolman-Oppenheimer-Volkoff (TOV) equations, supplemented by an assumed  equation of state, with continuous pressure $p\geq 0$  and a jump in the energy density $\rho$ at a pressure ``{\bf pjump}'', from an external relativistic matter state with $\rho=3p$  to an interior state with $p + \rho =\beta $, where $0<\beta<< 1 $.

 The dynamical gravastar formulation differs from that of Mazur and Mottola and the subsequent papers of Visser and Wiltshire \cite{visser} and Jampulski and Rezzola \cite{jamp} in several significant respects.  First, since \cite{adler1}, \cite{adler2} perform the entire analysis from the TOV equations, which require that the pressure $p$ must be continuous, whereas the energy density $\rho$ can have discontinuous jumps, they implement the Gliner equation of state by a jump to negative energy density with positive pressure.  This of course violates some (but as shown in \cite{adler2}, not all) of the classical energy conditions; however from a semiclassical quantum matter point of view, the regularized energy density is known {\it not} to obey positivity conditions \cite{wald}, \cite{Visser1}.  Second, whereas the original gravastar papers and more recent extensions such as \cite{jamp}  assume designated radii at which phase transitions take place,  in the dynamical gravastar model transitions follow dynamically from the equations of motion and the assumed equations of state, and the radii at which transitions occur are not specified in advance of solution of the dynamical equations.

Adler's second dynamical gravastar paper \cite{adler2} introduced two simplifications into the model constructed in his first paper \cite{adler1}, by setting the cosmological constant to zero (which has a negligible effect on the numerical results) and by replacing the smoothed sigmoidal function used for the energy density jump in \cite{adler1} with a step function jump (which changes numerical results but leaves qualitative features unaltered).  We shall only refer to results from the zero cosmological constant version of the model in this paper.  However, {\bf the   papers \cite{adler1} and \cite{adler2} do not address the question of whether the specific interior equation of state $p+\rho=\beta$ is needed to form a dynamical gravastar.  The purpose of the present paper is to show that a much wider class of interior equations of state suffices.}

\section{TOV equation system governing dynamical gravastars}

Calculations in general relativity begin by imposing coordinate conditions, to eliminate the ambiguities associated with covariance of the equations of relativity under general coordinate transformations \cite{weinberg}.  In calculations of stellar structure, the convenient choice of coordinates is spherical polar coordinates, in which the angular part of the line element has the form
\begin{equation}\label{ang}
ds^2=r^2 (d\theta^2+\sin^2\theta d\phi^2)+..,
\end{equation}
with the ellipsis denoting radial and time terms.  This choice of coordinates is used in both the texts on relativistic stars of Zeldovich and Novikov \cite{zeld} and of Camenzind \cite{camen}, and is the one they use to derive the TOV equations.  Once the coefficient of $d\theta^2+\sin^2\theta d\phi^2$ is fixed as $r^2$, no further coordinate transformation ambiguity is allowed, and the calculational results have a direct physical significance, unambiguously describing physics in the chosen coordinate system.  Changing to a different coordinate system can lead to a different description, even though the underlying physical process remains the same. Invariant quantities such as particle fluxes at infinity will be the same, whichever coordinate system is chosen to work with.  In comparing two different physical systems, such as dynamical gravastars with no horizon, and mathematical black holes with horizons, it is important to use the same choice of coordinates in describing both.  This is done consistently in this paper, where we  use the description of black holes by the Schwarzschild metric in spherical polar coordinates, corresponding to the coordinate choice used in the TOV equations.

The reason for reviewing this elementary material on coordinate fixing is to stress that since
{\bf the TOV equations are formulated within the framework of standard spherical polar coordinates,  the various functions appearing in them have no residual coordinate transformation freedom, and therefore describe physics in an unambiguous way, and are directly comparable to standard Schwarzschild metric functions in the same coordinate system.}

\subsection{The original form of the TOV equations and its rescaling covariance}

The TOV equations as given in \cite{adler2} for general density  $\rho$, followed in the second equality below  by their specialization to a relativistic gas with $\rho=3 p$,  are

\begin{align}\label{newTOV}
\frac{dm(r)}{d r}=&4\pi r^2\rho(r)= 12\pi r^2 p(r)~~~,\cr
\frac{dp(r)}{d r}=&-\frac{\rho(r)+p(r)}{2} \frac{d\nu(r)}{d r}= - 2 p(r) \frac{d\nu(r)}{d r} ~~~,\cr
\frac{d\nu(r)}{d r}=&\frac{N(r)}{1-2m(r)/r}~~~,\cr
N(r)=&(2/r^2)\big(m(r)+4\pi r^3 p(r)\big)~~~,\cr
\end{align}
where $p(r)$ and $\rho(r)$ are respectively the pressure and energy density, $m(r)$ is the volume integrated energy density within radius $r$,   and $\nu(r)=\log(g_{00}(r))$.   The  general form TOV equations become a closed system when  supplemented by an equation of state $\rho(p)$ giving the energy density in terms of the pressure. In
\cite{adler1} the equation of state used was a relativistic matter equation of state $\rho(p)=3p$ for $p\leq {\rm pjump}$, and $p + \rho(p) =\beta$, for $p > {\rm pjump}$.  In this paper, since we are focusing only on the external region of small pressure, with relativistic matter, we simply take $\rho(p)=3p$, as in the final equalities in Eq. \eqref{newTOV}.

It is useful in numerical work to use the following rescaling covariance of the TOV equations. Under the rescalings by a general parameter $L$,
\begin{align}\label{rescaling}
r \to &r/L~~~,\cr
m(r) \to & m(r/L)/ L~~~,\cr
\nu(r) \to & \nu(r/L)~~~,\cr
p(r) \to & p(r/L) L^2~~~,\cr
\rho(r) \to & \rho(r/L) L^2 ~~~.\cr
\end{align}
the TOV equation system of Eq. \eqref{newTOV} is form invariant.  This can be easily checked by direct substitution.  Rescaling covariance allows one to set the central pressure $p(0)$ to unity, as was done in \cite{adler1}; later on in this paper we will use it to set the initial radius to $r_0=50$ in the phase space diagram calculation in Fig. 6.  Either of these choices amounts to specifying the size  of the unit in which distances $r$ are measured.

\subsection{A rescaling invariant form of the exterior TOV equations}

The coupled equations for $p(r)$ and $m(r)$ given in Eq. \eqref{newTOV} can be recast in a form manifestly invariant under the scale transformation of Eq. \eqref{rescaling}.  Define new  quantities
\begin{align}\label{invs}
t=&\log{r} ~~~,\cr
dt =& dr/r ~~~,\cr
\alpha(t)=& m(r)/r~~~,\cr
\delta(t)=&4 \pi r^2 p(r)~~~.\cr
\end{align}
Scale transformation corresponds to shifting the origin of $t$, and the remaining three quantities $dt$, $\alpha(t)$, and $\delta(t)$ are invariants under this shift.  In terms of these variables, the TOV equations for $p$ and $m$ take the scale-invariant form
\begin{align}\label{scaleinvTOV}
\frac{d\alpha(t)}{dt}=& 3 \delta(t)-\alpha(t)~~~,\cr
\frac{d\delta(t)}{dt}=& -4 \delta(t) \frac{\delta(t)+2 \alpha(t)-1/2}{1-2\alpha(t)}~~~.\cr
\end{align}
Sample solutions obtained from the scale-invariant TOV equations are shown below in Fig. 4 and Fig. 5.\footnote{Figures Figs. 4--7 are computed using only the exterior TOV equations with appropriate initial conditions. The equations for the full model including the density jump of \cite{adler1}, \cite{adler2}  were used to calculate Figs. 1--3, which are qualitatively similar, but not identical, to Figs. 1, 6, and 9 of \cite{adler1}.   The latter were computed with different parameter values, using the original notebooks that include very small cosmological constant terms.   }

Remarkably,  these equations are what are called ``autonomous'' differential equations, in that no functions of the independent variable $t$ appear in the coefficients! This autonomous equation system describes a 2-dimensional flow,  a feature that may facilitate their  further mathematical analysis.  In this paper, we shall see below that the
scale-invariant equations are easier for the Mathematica integrator to follow, allowing a survey of a wider domain of initial values without incurring diagnostics, than was permitted by the original form of the TOV equations.

 \section{The gravastar ``simulated horizon'' defined;  it is a property of the TOV equation for relativistic matter}

\subsection{Simulated horizon in spherical polar coordinates}
 A typical result obtained from the dynamical gravastar model is shown in Fig. 1, where we have plotted the denominator function
 \begin{equation}\label{denomdef}
 D(r)\equiv 1-2m(r)/r
 \end{equation}
 appearing in the TOV equations of Eq. (2), with $m(r)$ the volume  integrated energy density contained within radius $r$.   This graph was plotted using the simplified model of \cite{adler2} in which the cosmological constant is zero, with parameters $\beta=.01$ and pjump=.95, where the units of distance for the computation were defined by taking the central pressure as $p(0)=1$ (which is always possible by using the rescaling covariance discussed in Sec. 3.)     {\bf We define the ``simulated horizon'' to be the location of the minimum of the denominator function $D(r)$, that is, the radius $r_{\rm min}$ where $dD(r)/dr|_{r=r_{\rm min}}=0$, with positive second derivative.} From Fig. 1  we see that the energy density jump at $p={\rm pjump}$, which occurs at a radius $r\simeq48.895$, occurs well inside the kink  structure around the simulated horizon, which lies at the radius $r\simeq 59.43754$. {\bf Hence the kink structure and generation of the simulated horizon are entirely a property of the TOV equations with relativistic matter and continuous pressure and density, when supplied appropriate initial values of the pressure and integrated mass density at the inner radius where the energy  density jump occurs.}  This is the viewpoint from which we will study the dynamical gravastar model in the present paper.  Thus, the kink structure in the external region will be the same for any evolution inside the jump radius that produces  the same initial  data at the jump radius.
  In particular, the results we obtain in this paper imply that the generation of a gravastar simulated horizon does not require the specific form of the interior region equation of state assumed in \cite{adler1}, \cite{adler2}.  The equation of state used in \cite{adler1}, \cite{adler2} gives a {\it sufficient condition} for the formation of a simulated horizon, or phrased another way, shows that the class or set of interior equations of state for which  a dynamical gravastar forms is not empty.    But the results we obtain in this paper show that the {\it necessary conditions} for the formation of a simulated horizon are much weaker.

\subsection{Extensions of the Simulated Horizon Analysis}

So far we have defined the simulated horizon in a specific choice of coordinates, the spherical polar coordinates used historically to derive the TOV equations.  However, our definition applies to more general coordinates. Consider the coordinate transformation from the original coordinates $r,\, t$ to new coordinates $r', \, t'$,
\begin{align}\label{transform}
r=&f(r')~~~,\cr
t=&t'~~~.\cr
\end{align}
Making this substitution in the TOV equations Eq. \eqref{newTOV}, we get in the new coordinate system, with $f'\equiv dr/dr'=df(r')/dr'$, and using $\rho(p)=3p$,
\begin{align}\label{newTOV1}
\frac{dm(r)}{d r'}=& 12\pi r^2 f' p(r)~~~,\cr
\frac{dp(r)}{d r'}=& - 2 p(r)f' \frac{N(r)}{D(r)} ~~~,\cr
N(r)=&(2/r^2)\big(m(r)+4\pi r^3 p(r)\big)~~~,\cr
D(r)=&D(f(r'))=1-2m(r)/r~~~.\cr
\end{align}
We see that in the new coordinates, the general form of the TOV equations is preserved, with a new denominator function $\hat D(r')=D(f(r'))$.

Letting $r'_{\rm min}$ be the image in the new coordinates of the simulated horizon originally calculated in the old coordinates, so that $r_{\rm min}=f(r'_{\rm min})$, we have
\begin{equation}\label{newhorizon}
d \hat D(r')/dr'|_{r'=r'_{\rm min}}=f' dD/dr|_{r=r_{\rm min}} =0~~~,
\end{equation}
and for the second derivative
\begin{align}\label{newcurvature}
&d^2 \hat D(r')/(dr')^2|_{r'=r'_{\rm min}}=d^2D(r)/(dr)^2|_{r=r_{\rm min}} (f')^2|_{r'=r'_{\rm min}} + dD(r)/dr|_{r=r_{\rm min}}  df'/dr'|_{r'=r'_{\rm min}} \cr
&=d^2D(r)/(dr)^2|_{r=r_{\rm min}} (f')^2|_{r'=r'_{\rm min}}>0~~~,\cr
\end{align}
showing that the conditions defining the simulated horizon are reproduced in the new coordinate system.\footnote{From a global perspective, this is not surprising.  A finite segment of a convex upwards curve around a minimum retains  this property irrespective of the parametrization of the variable running along the curve.}    The mapping
of Eq. \eqref{transform} is general enough to include the transition from spherical polar coordinates to isotropic coordinates.     Thus the simulated horizon concept is general enough to apply to both types coordinate systems in which  the TOV equations are used for calculation of relativistic stars \cite{isotropic}.

\subsection{ Coordinate Independent Property Associated with the Simulated Horizon, and Comparison with the Black Hole Event Horizon}

Just to check that our use of the term ``simulated horizon'' is appropriate, in  Fig. 2 we plot $g_{00}(r)=e^{\nu(r)}$ from the dynamical gravastar calculation used in Fig. 1. This is nearly indistinguishable from the exterior Schwarzschild geometry for a black hole of mass $M
 \simeq 29.72$ and Schwarzschild radius $r\simeq 59.44$.   However, the logarithmic plot in Fig. 3 shows that within the simulated horizon, $g_{00}$ for the dynamical gravastar remains always positive, but becomes very small, which means that  {\bf objects that enter the simulated horizon can get back out. The fraction of  objects that enter the gravastar at infinity and eventually exit the gravastar at infinity is a coordinate-independent quantity associated with the simulated horizon.}\footnote{ In asserting coordinate independence here, we are restricting to the class of asymptotically flat coordinates, which includes spherical polar and isotropic coordinates.} It will be nonzero as long as long as the TOV evolution maintains $2\alpha(t)=2m(r)/r <1$ so that the TOV denominator $D$ is always strictly positive and $\nu(r)$ is nonsingular.    This contrasts with the behavior at the horizon of a Schwarzschild black hole:
  $g_{00}$ for a free space Schwarzschild solution with the same mass would become negative inside $r\simeq59.44$,  following the curve $g_{00}\simeq 1-59.44/r$ that is the continuous leftward extrapolation of the segment of the curve in Fig. 2 on the right of $r \simeq 59.44$, and as is well known, objects that enter a Schwarzschild horizon cannot get back out.\footnote{An event horizon in a stationary asymptotically flat spacetime can be characterized globally as a Killing horizon for a Killing vector field. However, global concepts are not useful in numerical relativity, as well as for time-evolving black holes, because one has no access to the global spacetime.  This is well discussed in two review articles \cite{farioni}, \cite{krishnan}  which show why in numerical relativity one falls back on quasi-local horizon criteria, especially the concept of the ``apparent horizon'', based on the presence or absence of trapped surfaces. The  simulated horizon introduced in Sec. 3A of the present paper is another example of a quasi-local, rather than a global horizon concept.}

\begin{figure}[]
\begin{centering}
\includegraphics[natwidth=\textwidth,natheight=300,scale=0.8]{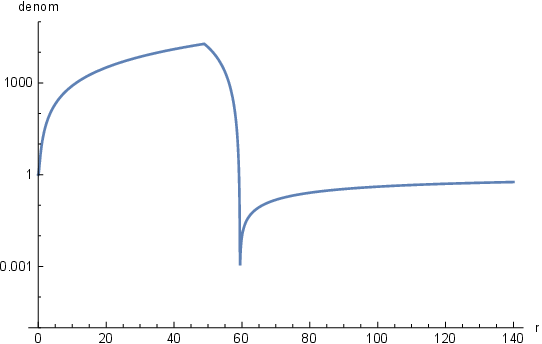}
\caption{Plot of $D(r)$ for the TOV.01 notebook.  The energy density jump at $r=48.895$ is apparent from the slope discontinuity; above this radius the functions entering the TOV equations are continuous, as can be verified by zooming in on the apparent cusp near $r\simeq 59.43754$ with a finer plotting scale. For further discussion of smoothness of the exterior region solutions, see Sec. IV.}
\end{centering}
\end{figure}

\begin{figure}[]
\begin{centering}
\includegraphics[natwidth=\textwidth,natheight=300,scale=0.8]{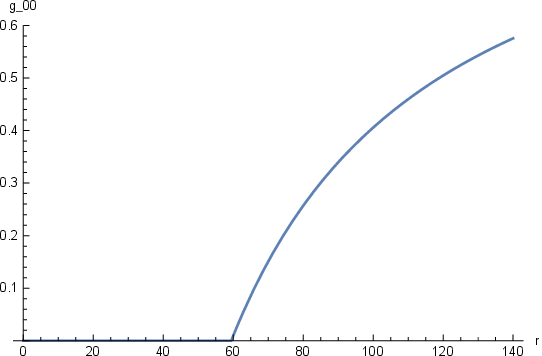}
\caption{Plot of $g_{00}$ for the same parameters used in Fig. 1.  Outside the radius $r\simeq 59.44$, this plot is nearly indistinguishable from a plot of the exterior Schwarzschild geometry for a black hole of mass $M=29.721$.   }
\end{centering}
\end{figure}

\begin{figure}[]
\begin{centering}
\includegraphics[natwidth=\textwidth,natheight=300,scale=0.8]{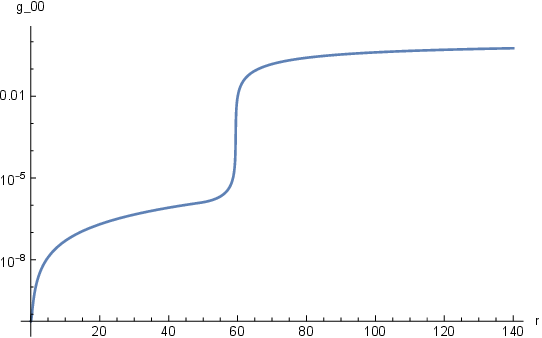}
\caption{Log plot of $g_{00}$ for the same parameters used in Fig. 1. This shows that $g_{00}$ is always positive within the simulated horizon, but becomes very small.}
\end{centering}
\end{figure}
\begin{figure}[]
\begin{centering}
\includegraphics[natwidth=\textwidth,natheight=300,scale=0.8]{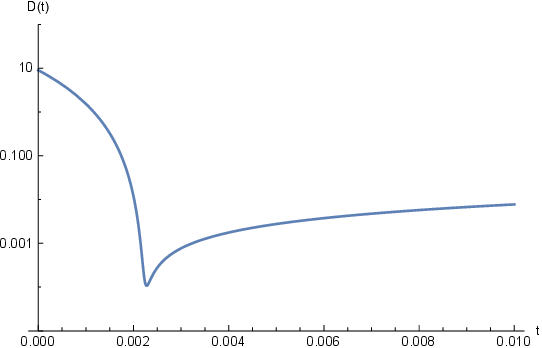}
\caption{Kink solution obtained from the scale invariant TOV equations, with initial values $\alpha_0=-4, \, \delta_0=2,000$.    }
\end{centering}
\end{figure}

\begin{figure}[]
\begin{centering}
\includegraphics[natwidth=\textwidth,natheight=300,scale=0.8]{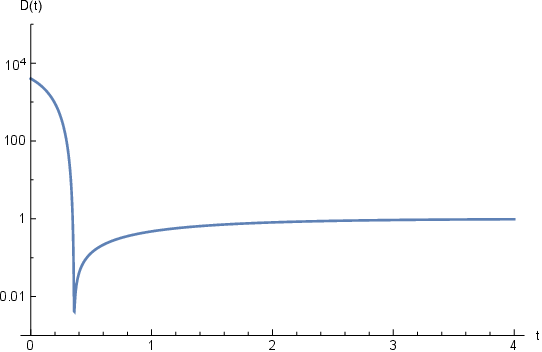}
\caption{Kink solution obtained from the scale-invariant TOV equations, with initial values $\alpha_0=-2,000, \, \delta_0=2,000$. A much deeper and sharper cusp is obtained than with the initial values used in Fig. 4.}
\end{centering}
\end{figure}

\section{Phase diagram  in the initial pressure -- initial mass plane}

\subsection{Survey using the orginal form of the TOV equations in Eq. \eqref{newTOV}}

The kink  behavior leading to a simulated horizon exhibited in the previous section is typical of a wide range of parameter values.  To see this, let us numerically study in more detail the differential equation system of Eq. \eqref{newTOV} for $p(r)$ and $m(r)$ .   We use the rescaling covariance of Eq. \eqref{rescaling} to rescale the initial radius $r_0$ to a standard value $r_0=50$, chosen to approximate the energy density jump radius $r=48.895$ of Fig. 1, so that a solution close to the one shown there is included in our survey.     The remaining parameters of the model are then the initial values $p_0\equiv p(r_0)$ and $m_0\equiv m(r_0)$.  We ran the integration of Eqs. \eqref{newTOV} for a wide range of initial values $p_0$, ranging from $p_0=.01$ to $p_0=10^4$, and for each determined the largest value of $m_0$ for which a kink solution resembling Fig. 1 is obtained, as tabulated in Table 1.  In the final two columns of Table 1 we  give the corresponding initial values of the scale-invariant variables $\alpha, \, \delta$  defined in Eq. \eqref{invs}, which we will apply in the next subsection to surveys of the phase space using the scale-invariant form of the TOV equations.

For larger values of $m_0$ than  the limiting values shown in Table 1, the Mathematica integrator fails when applied to Eq. \eqref{newTOV}.\footnote{Specifically, the integrator NDSolve returns an error message of a ``stiff system'', where we have used parameters PrecisionGoal and AccuracyGoal of 13 (the maximum attainable on our hardware) and MaxSteps of $10^{10}$.}    The results summarized in Table 1 can be plotted as a ``phase diagram'' in the $p_0$--$m_0$ plane as shown in Fig. 6, with points in the shaded region giving a kink solution, and points outside the shaded region giving an integrator error message.  This demonstrates clearly that the horizon-like behavior of Figs. 1 and 2 is not limited to the specific interior equation of state assumed in \cite{adler1}, \cite{adler2},  but is characteristic of the TOV equations with relativistic matter for a wide range of initial values $p_0$, $m_0$ at the inner radius $r_0$ where the integration is started.\footnote{Further plots of kinks, and related Tables,  using the original form of the TOV equations can be found in arXiv:2309.13380v7. These  show the trend of kink shapes as the parameters $p_0$ and $m_0$ are varied}

\subsection{Survey using the scale-invariant form of the TOV equations given in Eq. \eqref{scaleinvTOV}}

The scale-invariant form of the TOV equations  for relativistic matter given in Eqs. \eqref{scaleinvTOV} was discovered  by SLA  several months after the phase diagram of Fig. 6 was plotted.  Using Eqs. \eqref{scaleinvTOV}, we found gratifyingly that the phase diagram could be considerably enlarged.  Specifically, starting from an initial $t$ value chosen conveniently as $t_0=0$, for all $\delta_0$ values in Table 1, $\alpha_0$ could be enlarged to a value close to $0.5$ before the Mathematica integrator fails.\footnote{Specifically, for all $\delta_0$ in Table 1, $\alpha_0$ could be enlarged to at least $0.49$ before a ``stiff system'' diagnostic appears, and for all $\delta_0$ values values at and above $993371$, $\alpha_0$ could be enlarged to at least $0.4999$ before a ``stiff system'' diagnostic appears.}  In all cases calculated with $\alpha_0<1/2$ and $\delta_0>0$,  a kink solution is obtained, provided only that $3 \delta_0-\alpha_0>0$,  which specifies that the initial value of the derivative $d\alpha/dt$ is positive.     Since $\delta_0 >0$, this  condition is automatically satisfied for negative values of $\alpha_0$, for which a kink is always obtained.  Thus, all of our numerical results are consistent with the following conjecture:

{\bf Conjecture:  Provided that $m_0/r_0=\alpha_0<1/2$ and $4\pi p_0 r_0^2 =\delta_0>0$, and $3\delta_0-\alpha_0>0$, the TOV equations for relativistic matter  yield a kink solution or ``simulated horizon''.}  Remark:  The condition $m_0/r_0=\alpha_0<1/2$  is clearly a necessary condition; the condition $3 \delta_0-\alpha_0>0$ excludes the case when $D$ develops a local maximum, and could possibly be weakened.

We leave proving this conjecture as a  challenge for the mathematical community!  Another important mathematical question is to try to get approximate formulas for the depth and width of the kink minimum as a function of the inner radius boundary values.

The qualitative structure of the solution in the rescaling invariant variables depends on the relative size of the initial values $\delta_0$ and $\alpha_0$.
For $\delta_0+2\alpha_0-1/2>0$, the variable $\delta$ initially decreases with increasing $t$,  while for $\delta_0+2\alpha_0-1/2<0$, the variable $\delta$ initially increases  with increasing $t$, and then turns around and decreases.  With $\delta$ decreasing, and $\alpha$ increasing, eventually the right hand side of the equation $d\alpha(t)/dt=3\delta(t)-\alpha(t)$ changes from positive to negative, giving the kink minimum of $D(t)=1-2\alpha(t)$ characterizing the ``simulated horizon''.\footnote{Note that since $D(r)=1-2 \alpha(t)$, the condition $d\alpha/dt=0$ specifies the same simulated horizon radius as the original condition $dD(r)/dr=0$. }                 This is illustrated in Fig. 7, drawn for the same parameter values as used for Fig. 4.

From the scale-invariant form of the TOV equations, one can derive an approximate form of $\alpha(t)$ and $\delta(t)$ in the neighborhood of a very deep and sharp cusp minimum in $D(t)$.  Since the equations are $t$ translation invariant,  with no loss of generality one can now take the $t$ value of the minimum to be $t=0$.  Then one finds,
\begin{align}\label{dipform}
\alpha(t) \simeq & 1/2- \theta_0 - \theta_2 t^2+O(t^3)~~~,\cr
\delta(t) \simeq &  1/6 -(2/9)\rm{ArcTan}\left( (\theta_2/\theta_1)^{1/2} t  \right) / (\theta_2 \theta_1)^{1/2}+O(\theta_1,\theta_2, t^2)~~~,\cr
\end{align}
where $ \rm{ArcTan}(0)=0 $ and  $\rm{ArcTan}(\infty)=\pi/2 $, and with $\theta_0<<1$ and $\theta_2<<1$ parameters determining the depth and width of the minimum.   These parameters are fixed by  integration of the TOV equations out from the inner radius where initial data is specified.\footnote{For an ArcTan fit to the behavior of $\nu(r)$ near the simulated horizon in Fig. 1, see Sec. IV of arXiv:2309.13380v7.}

\begin{figure}[]
\begin{centering}
\includegraphics[natwidth=\textwidth,natheight=300,scale=0.8]{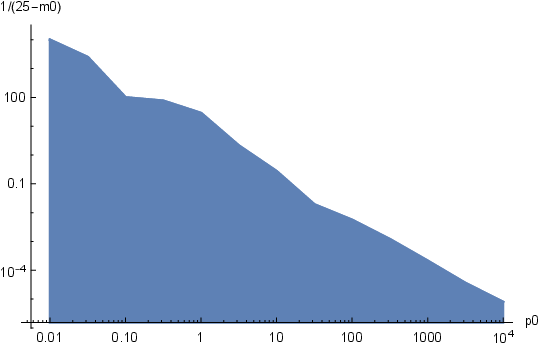}
\caption{``Phase diagram'' in the $p_0$,  $(25-m_0)^{-1}$ plane, where $p_0=p(r_0)$ is the initial pressure and and $m_0=m(r_0)$ is the initial mass. The diagonal boundary corresponds to the largest value of $m_0$ for which the Mathematica notebook gives a kink solution from the original form of the TOV equations.  For values of $m_0$ corresponding to the region above this boundary the program gives an  error diagnostic.  For points within the shaded region, a kink solution is obtained.  The scale-invariant TOV equations shift the diagonal upper boundary upwards.   }
\end{centering}
\end{figure}

\begin{figure}[]
\begin{centering}
\includegraphics[natwidth=\textwidth,natheight=300,scale=0.8]{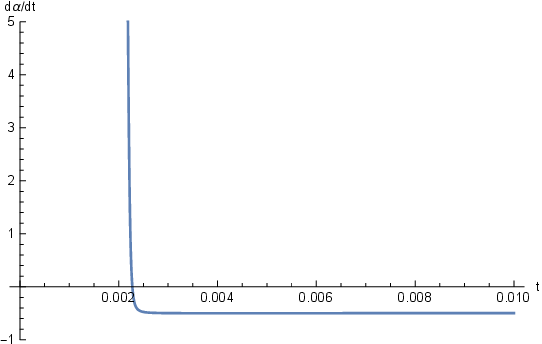}
\caption{Plot of $d\alpha/dt$ showing the zero crossing that gives the minimum of $D(t)$ in Fig. 4   }
\end{centering}
\end{figure}

Also from the scale-invariant form of the TOV equations, significant results on smoothness of the solutions follow.  Since the right-hand side of Eqs. \eqref{scaleinvTOV} is smooth except for a simple pole in $\alpha(t)$ at $\alpha=1/2$, it is $C^\infty$, in other words it has continuous derivatives of all orders with respect to $\alpha$ and $\delta$,  in the domain $ \Delta \equiv \{\alpha<1/2,\,\delta>0\}$.  This implies \cite{mathencyc} that $\alpha(t)$ and $\delta(t)$ are $C^\infty$ functions of $t$ for $\alpha,\, \delta$ values residing in the domain $\Delta$.  This can be proved  by multiple differentiation of Eqs. \eqref{scaleinvTOV} with respect to $t$. Since the order $p+1$ derivative with respect to $t$ of $\alpha,\, \delta$ on the left-hand side is expressed as a linear combination of order $p$ or lower derivatives of $\alpha,\, \delta$ on the right-hand side, continuity of the left-hand side follows from continuity on the right, yielding a proof by mathematical induction.  This implies that the cusp at the simulated horizon, no matter how sharp it looks in a plot, is actually  smooth to all orders of derivatives!  As pointed out in Sec. II of Ref. \cite{adler2}, smoothness of the kink solution plays an important role in gravastar evasion of certain no-go results \cite{nogo1}--\cite{nogo3} for exotic compact objects.

\section{Novel results of this paper}

This paper is the culmination of an investigation of exotic compact objects with dynamical underpinnings  begun in \cite{adler1} and \cite{adler2}.  The paper \cite{adler1} showed that a sufficient  condition to get a fully dynamical gravastar as a solution to the TOV equations is to take an interior Gliner equation of state $p+\rho=\beta$ with $\beta$ a small positive constant. This equation of state is agonostic with respect to whether the pressure $p$ or the energy density $\rho$ should under go a jump to negative values, arising from a phase transition.  But the TOV equations (as noted in \cite{adler1}) require pressure to be continuous, while allowing the energy density to jump, consistent with the fact  that in quantum theory the energy density is no longer necessarily positive. The paper \cite{adler2} extended the results of \cite{adler1} in a number of ways, including showing that the dynamical gravastar model can evade various ``no-go'' arguments for exotic compact objects, and showing that this model still obeys the null and strong energy conditions that are used to prove black hole singularity theorems.  However, the interior equation of state assumed in \cite{adler2} was the same as used in \cite{adler1}.

The purpose of the present paper is to study in more detail the mechanism by which the dynamical gravastar model produces a horizon-like structure, with the metric component $g_{00}$  remaining strictly positive.  We focus on the exterior region solution, above the energy density  discontinuity, assuming that the final state of matter evolution before collapse is a fully relativistic equation of state  $\rho=3p$, which is described in the classic text of Zeldovich and Novikov \cite{zeld} as the limiting case of relativistic matter equations of state.  We find that the creation of what we term a
``simulated horizon'' is entirely a property of the external region TOV equation with an appropriate inner boundary condition, and achieve novel results that impact the following three areas.

\begin{itemize}
\item {\bf Theory of the TOV equation} Although Buchdahl \cite{buch} has given general conditions for evasion of a horizon, his inequalities do not give the shapes of horizon-less solutions of the Einstein equations when his conditions are violated.    For this one has to turn to numerical and graphical studies of the TOV equations. The TOV equations have been known for close to a century, and are widely applied to numerical calculations in relativistic stellar dynamics.  But what has {\it not} been done before is to analyze what happens when the  TOV equations are solved with a negative energy density inner boundary condition; stellar structure and neutron star applications assume a positive inner energy density.  As developed in this paper, the surprising new conclusion emerges, that with a negative inner energy density the TOV equations develop a ``simulated horizon'', in which $g_{00}$ becomes exponentially small, but never crosses over to negative values.  This gives the detailed mechanism by which the dynamical gravastar solution of \cite{adler1} is obtained, in a wide class of interior equations of state.  The very specific  condition $p+\rho=\beta$ used in \cite{adler1} and \cite{adler2} is generalized to much weaker  conditions plotted in our phase diagram surveys.\footnote{The specific equation of state chosen  does play an important role in determining the radius at which the simulated horizon appears, and in determining how deep and sharp is the associated cusp or kink, since it determines the values of $m(r)$ and $p(r)$, or equivalently $\alpha(t)$ and $\delta(t)$, at the radius where the density jumps.}  In addition, we obtain the novel result that the relativistic matter TOV equation can be rewritten as a pair of coupled autonomous differential equations, implying through the known mathematical theory of such systems that the sharp kinks observed in the plots
    are in fact $C^\infty$, a result of relevance for the argument in \cite{adler2} that dynamical gravastars can evade ``no-go'' claims for ECOs.
\item{\bf Improvement on the original Mazur-Mottola model and its descendents} The Mazur-Mottola papers introduced an ingenious idea, but left a key question unanswered:  What mechanism fixes the pre-assigned radii at which structural jumps occur?  This lacuna is filled by the analysis of the present paper, which shows that dynamically occurring jumps are a natural by-product of the relativistic matter TOV equation with a negative energy density interior boundary condition.
\item{\bf Implications for astrophysics}  Whether observed ``black holes'' have true horizons, through which nothing can escape, or only approximate horizons, where particle fluxes can leak out, will impact the modeling of astrophysical phenomena.  As one example, both Adler \cite{adler3}, \cite{adler4} and Silk et. al. \cite{silk} have proposed models for galaxy formation nucleated by a supermassive central black hole, with outward flowing particle ``winds'' playing an important role.  The natural occurrence of ``leaky'' simulated horizons proposed in the present article is relevant for the further development of these ideas.  As a second example, in principle there can be large time delays between entry of an object to a gravastar and its eventual exit.  If astrophysical black holes are gravastars, this could explain the recent report by Cendes et al. \cite{cendes} that many black hole tidal disruption events have subsequent radio emissions delayed by hundreds to thousands of days.
\end{itemize}

\section{Acknowledgements}

We thank Ahmed ElBanna and Yuqui Li for advice on how to incorporate  Mathematica utilities for maximization and minimization into our computations.  Dr. ElBanna has placed an annotated version of the Mathematica notebook used for Fig. 6 and Table 1 online at the following URL:
https://community.wolfram.com/groups/-/m/t/3022687~~~.  SLA also wishes to thank Thomas Spencer and Michael Weinstein for a stimulating conversation about issues addressed in this paper, and  to acknowledge the hospitality of Clare Hall in Cambridge, U.K. during revisions of this paper.

\begin{table} [ht]
\caption{Boundary points used to plot Fig. 6.  When recalculated using the scale-invariant form of the TOV equations, the $\alpha_0$ values in the third column get pushed up to close to $0.5$, corresponding to an upwards relocation of the diagonal upper boundary in Fig. 6. }
\centering
\begin{tabular}{c c c c}
\hline\hline
$p_0$&$m_0$&$\alpha_0$&$\delta_0$\\
\hline
~~~~0.01~~~~& 24.9998999& .499998    & 314.159      \\
~~~~.03162~~~~&24.9996004& .499992    &993.37      \\
~~~~0.1~~~~~& 24.9899999 & .499800    & 3141.59           \\
~~~~.3162~~~~&24.9869469&.499739     & 9933.71          \\
~~~~~1~~~~~~& 24.9651 & .499302    & 31415.9          \\
~~~~3.162~~~&24.5310&.49012     & 99337.1           \\
~~~~10~~~~~~& 21.366 &.42732     &314159         \\
~~~~31.62~~~~&-26.75&-.535     & 993371         \\
~~~~100~~~~~& -153 &-3.06     &3141590          \\
~~~~316.2~~~~&-806&-16.12     & 9933708         \\
~~~~1000~~~~~& -4,640& -92.8   & 31415900        \\
~~~~3162~~~~&-27,600& -63.24    & 99337080         \\
~~~~10000~~~&-126,000&-2520     & 314159000         \\
\hline\hline
\end{tabular}
\label{tab1}
\end{table}

\end{document}